\documentclass[fleqn,twoside]{article}
\usepackage{espcrc2}
\newif\ifpdf
\ifx\pdfoutput\undefined\pdffalse\else\pdfoutput=1\pdftrue\fi
\ifpdf\pdfcompresslevel=9\fi 
\def\epspdffile#1{\leavevmode\ifpdf\epsffile{#1.pdf}\else\epsffile{#1.eps}\fi}
\usepackage{epsfig}
\def\min{\mathop{\rm min}}
\def\ulp{\mathop{\rm ulp}}
\def\max{\mathop{\rm max}}
\def\det{\mathop{\rm det}}
\def\tr{\mathop{\rm tr}}
\def\ln{\mathop{\rm ln}}

\def\spf{\mathop{S_{\hbox{\rm\tiny PF}}}}
\def\nf{n_{\rm f}}
\def\dt{\delta t}
\def\m{\mathcal{M}}
\def\hpf{H}				
\def\hmd{\bar H}			
\def\rpf{r}				
\def\rmd{\bar r}			
\def\dpf{\Delta}			
\def\dmd{\bar\Delta}			
\def\Pacc{P_{\hbox{\rm\tiny ACC}}}
\def\half{\frac12}

\begin{document}

\title{The RHMC Algorithm for 2 Flavours of Dynamical Staggered Fermions}

\author{M. A. Clark and A. D. Kennedy\address{School of Physics, University of
   Edinburgh, King's Buildings, Edinburgh, EH9 3JZ, United Kingdom}}

\begin{abstract}
  {\noindent We describe an implementation of the Rational Hybrid Monte Carlo
  (RHMC) algorithm for dynamical computations with two flavours of staggered
  quarks. We discuss several variants of the method, the performance and
  possible sources of error for each of them, and we compare the performance
  and results to the inexact \(R\) algorithm.\parfillskip=0pt\par}
\end{abstract}

\maketitle

\section{Introduction}

Computations with two flavours of dynamical staggered quarks are quite popular
at present. There are a number of possible problems with such calculations such
as flavour symmetry breaking and non-locality of the square-root of the
four-flavour action. In this investigation we shall ignore these and consider
only the possible errors introduced through algorithmic approximations.

We propose the use of the Rational Hybrid Monte Carlo (RHMC) algorithm
\cite{kennedy98a}. This method is stochastically exact, in the sense that it is
free from molecular dynamics (MD) stepsize errors. It is comparable to the
usual \(R\) algorithm \cite{gottlieb87a} in performance, but without the need
for extrapolation in the MD stepsize~\(\dt\).

\section{Two Flavour Algorithms}

All Hybrid Molecular Dynamics (HMD) algorithms have the same underlying
structure: a fictitious momentum field is introduced, and the gauge field is
integrated along classical trajectories in fictitious time, interleaved with
refreshment of the momenta from a Gaussian heatbath. When integrating
Hamilton's equations the evaluation of the fermionic contribution to the force
acting on the gauge fields is the costliest part of generating full QCD gauge
field configurations. Most algorithmic developments are techniques to calculate
the fermionic force more efficiently. 

The desired probability distribution for the gauge fields \(U\) with \(\nf\)
flavours of staggered fermions is
\begin{equation}
  P(U) \propto e^{-S_G(U)} \det\bigl[\m(U)\bigr]^{\nf/4}, \label{eq:pu}
\end{equation}
where \(\m\) is the staggered fermion kernel and \(S_G\) is the gauge action.
Thus for \(\nf=2\) flavours of fermion we require the square root of the
fermion determinant. Choosing a suitable normalisation the spectrum of the
staggered fermion kernel \(\m\) is contained in the interval
\([\varepsilon,1]\), where \(\varepsilon = \left(1+\frac{16} {m^2}\right)^{-1}
= \frac{m^2}{16} + O(m^4)\).

\subsection{The $R$ algorithm}

The identity \(\det\m^{\nf/4}=\exp\tr\ln\m^{\nf/4}\) allows us to express the
determinant as a term in the action, and the number of flavours just appears
as a factor in front of this term, \(\spf=-\frac\nf4\tr\ln\m\). In the \(R\)
algorithm \cite{gottlieb87a} the evaluation of the force corresponding to this
trace is performed stochastically, as computing it exactly would be
prohibitively expensive. To do this without introducing \(O(\dt)\) errors we
introduce an auxilliary field that is evaluated at time \((1-\frac\nf4)
\frac\dt2\) along each integration step. For two flavours this breaks
time-reversal invariance, violates Liouville's theorem, and leads to an
irreversible and non area-preserving algorithm: as such, it cannot be made
exact by the inclusion of a Metropolis step. A detailed analysis of the errors
in the probability distribution produced by this algorithm was presented
in~\cite{Clark:2002vz}.

\section{Rational Hybrid Molecular Dynamics}

\begin{figure}[t]
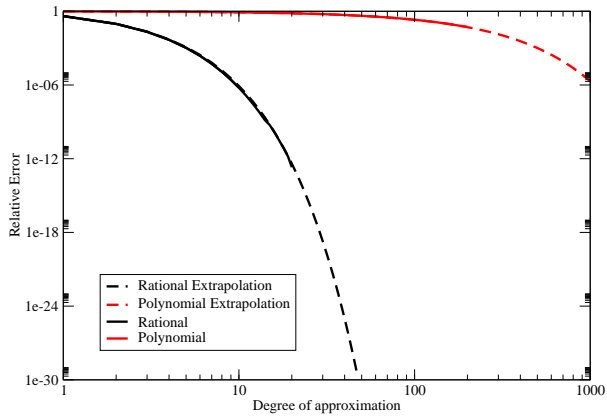

  \epsfxsize=0.5\textwidth
  \centerline{{\epspdffile{degree}}}
  \vskip-6ex
  \caption[Rational Error]{Comparison of minimax errors for optimal rational
    and polynomial approximating functions to \(x^{-1/2}\) over the range
    [0.00003,1] (corresponding to staggered mass parameter \(m=0.025\)) as a
    function of approximation degree.}
  \label{fig:rat-error}
\end{figure}

RHMD \cite{kennedy98a} uses a uses a rational approximation to fractional
powers \(\alpha\) of a matrix. This is analogous to the use of polynomial
approximations introduced in \cite{forcrand:1996,jansen97a,luescher94a}, but
optimal (Chebyshev) rational approximations give a much closer approximation
over a given interval than the corresponding polynomials of similar degree
(Fig.~\ref{fig:rat-error}). The best choice for this rational approximation
would seem to be given by a relative minimax approximation: this has the
property that it is guaranteed to deviate by at most an amount \(\dpf\)
relative to the correct value over a given interval, \(\max_{\varepsilon\leq
x\leq1} \left|1 - x^\alpha\rpf(x) \right| = \dpf\). The errors are of the same
kind as those introduced by the use of floating-point arithmetic. The maximum
error \(\dpf\) falls exponentially with the degree of the rational
approximation used.

We approximate the determinant in~(\ref{eq:pu}) as a Gaussian integral over a
bosonic (pseudofermion) field \(\phi\), \[\det\m^{\nf/4} \approx
\det\rpf(\m)^{-2} \propto\int d\phi d\phi^\dagger\, e^{-\phi^\dagger\rpf(\m)^2
\phi},\] where \(\|1 - \rpf(\m)\, \m^{\nf/8}\| \leq\dpf\) in the spectral norm.
The RHMD algorithm proceeds as follows
\begin{itemize}
\item A momentum refreshment heatbath using Gaussian noise: \(P(\pi)\propto
  e^{-\half|\pi|^2}\).
\item A pseudofermion heatbath using Gaussian noise: \(\phi = \left[\rpf
  \left(\m(U)\right)\right]^{-1} \xi\), where \(P(\xi) \propto
  e^{-\half|\xi|^2}\). Note that the inverse \(1/\rpf(x)\) is itself a rational
  function. 
\item An MD trajectory consisting of \(\tau/\dt\) steps with Hamiltonian \(\hmd
= \half |\pi|^2 + S_G + \phi^\dagger \rmd(\m)\phi\) with \(\|1 - \rmd(\m)\,
\m^{\nf/4} \| \leq\dmd\) in the spectral norm.
\end{itemize}
This leads to an algorithm which has finite stepsize errors of \(O(\dt^2)\) and
errors of \(O(\dmd)\) and \(O(\dpf)\) incurred from the use of rational
approximations.

Rational functions can be expressed as a product or as a partial fraction
expansion, \[\rmd(x) = \bar c_0 \prod_{i=1}^d \frac{(x-\bar\gamma_i)} {(x-
\bar\beta_i)} = \bar\alpha_0 + \sum_{i=1}^d \frac{\bar\alpha_i}{x-
\bar\beta_i}.\]

In partial fraction form \cite{Neuberger:1998my} the pseudofermionic force
takes the form \[\frac{\partial\hmd}{\partial U} = \phi^\dagger
\frac{\partial\rmd(\m)} {\partial U} \phi = - \sum_{i=1}^d \bar\alpha_i
\chi_i^\dagger \frac{\partial \m} {\partial U}\chi_i\] where \(\chi_i =
(\m-\bar\beta_i)^{-1} \phi\). A multishift solver \cite{Frommer:1995ik} can be
used to compute all the \(\chi_i\) in a common Krylov space. The computational
cost of generating the appropriate Krylov space depends upon the smallest
shift, and the only extra cost is that of updating the extra \(d-1\) solution
vectors.

\section{Rational Hybrid Monte Carlo}

The RHMC algorithm is similar to RHMD but with the addition of a Metropolis
accept/reject step. The acceptance probability for this is given by
\begin{equation}
  \Pacc = \min\left(1, e^{\delta\hpf} \frac{\det{\m'}^{\nf/4}\rpf(\m')^2}
    {\det\m^{\nf/4}\rpf(\m)^2} \right)
  \label{eq:probrat}
\end{equation}
where \(\hpf = \half|\pi|^2 + S_G + \phi^\dagger \rpf(\m)^2 \phi\).

If we use a high enough degree rational approximation \(\rpf\) such that
\(\dpf\leq1\ulp\) (unit of least precision for the floating point arithmetic
used) in the computation of the pseudofermion heatbath and of \(\delta\hpf\) we
can ignore the explicit determinants in (\ref{eq:probrat}) without introducing
any systematic errors beyond the ever-present rounding errors. In practice we
find that it is sufficiently cheap to use the same (machine accuracy) rational
approximation for the MD integration as well (\(\dmd=\dpf\)), although this is
not logically required.

\section{Noisy Rational Hybrid Monte Carlo}

RHMCN \cite{kennedy98a} allows the use of a lower degree rational approximation
\(\rpf\) while keeping the algorithm exact. The algorithm replaces the
Metropolis step with a Kennedy-Kuti (KK) \cite{kennedy85e} noisy accept/reject
step. A stochastic summation is used to estimate the determinant ratio
in~(\ref{eq:probrat}). The acceptance probability is defined as \(\Pacc =
\lambda_\pm + \lambda_\mp Q(U,U')\) for \(U>U'\) and \(U<U'\) respectively,
where \(Q(U,U')\) is an unbiased (noisy) estimator of the determinant ratio
occurring in~(\ref{eq:probrat}) and \(\lambda_\pm\) are parameters used to
ensure the resultant probability distribution lies in the range \([0,1]\). When
\(\lambda_+ = \lambda_- = \lambda\) the average acceptance rate is \(\langle
P\rangle=2\lambda\).

\section{Comparison of RHMC and $R$}

The use of multishift solvers in the implementation of the rational algorithms
results in a lower computational cost than might otherwise be expected. On a
\(16^3\times32\) lattice with \(\beta=5.26\), \(m=0.01\), and \(\dt=0.01\) the
time to perform one trajectory of length \(\tau=0.5\) on a single node
Pentium~4 processor with no assembler optimisations is 274 minutes for the
\(R\) algorithm. RHMC takes 318 minutes when a degree~10 rational function is
used. Although a single \(R\) trajectory is faster, RHMC compares very
favourably when it is remembered that an \(O(\dt^2)\) extrapolation ought to be
carried out when using the inexact \(R\) algorithm.

The computational cost of using RHMCN is unfavourable compared with RHMC.
Although a lower degree rational approximation can be used this does not give a
large benefit because of the efficacy of the multishift solver. To ensure
negligible probability violations occur when \(\Delta>1\ulp\) the KK acceptance
test becomes increasingly expensive for large volumes and small quark masses.
The cost of performing a single trajectory using the parameters given above
with a degree 6 rational approximation and typical stochastic summation
parameters is 385 minutes. We found that setting \(\lambda\) to give \(\langle
P\rangle=70\%\) gives as large an acceptance rate as feasible without incurring
excessive violations, but for large \(V\) and small \(m\) it was found
necessary to reduce this to about \(50\%\).

\section{Conclusion}

We have found that it is easy and cheap to compute rational powers of the
staggered fermion kernel to within machine accuracy using Chebyshev rational
approximations expressed as partial fractions, and applied using a multishift
solver. This form of the RHMC algorithm thus enables the exact HMC algorithm to
be extended to the case of an arbitrary number of flavours. The ability to use
Krylov space solvers makes RHMC faster than the PHMC
algorithm~\cite{forcrand:1996,jansen97a}. In terms of computational cost there
is very little to chose between the \(R\) and RHMC algorithms. Since this seems
to be the only possible advantage of \(R\) over RHMC, we conclude that there is
no reason for the continued use of the \(R\) algorithm.

\section*{Acknowledgments}

We would like to thank Robert Edwards, Ivan Horv\'ath, B\'alint Jo\'o, and
Stefan Sint for discussions.


\end{document}